\documentstyle[psfig]{mn}

\newif\ifAMStwofonts

\ifoldfss

  \ifCUPmtlplainloaded \else

    \NewTextAlphabet{textbfit} {cmbxti10} {}

    \NewTextAlphabet{textbfss} {cmssbx10} {}

    \NewMathAlphabet{mathbfit} {cmbxti10} {} 

    \NewMathAlphabet{mathbfss} {cmssbx10} {} 

  \fi

  \ifAMStwofonts

    \ifCUPmtlplainloaded \else

      \NewSymbolFont{upmath} {eurm10}

      \NewSymbolFont{AMSa} {msam10}

      \NewMathSymbol{\upi}     {0}{upmath}{19}

      \NewMathSymbol{\umu}     {0}{upmath}{16}

      \NewMathSymbol{\upartial}{0}{upmath}{40}

      \NewMathSymbol{\leqslant}{3}{AMSa}{36}

      \NewMathSymbol{\geqslant}{3}{AMSa}{3E}

      \let\leq=\leqslant 

      \let\geq=\geqslant 

    \fi

  \fi

\fi 

\ifnfssone

  \newmathalphabet{\mathit}

  \addtoversion{normal}{\mathit}{cmr}{m}{it}

  \addtoversion{bold}{\mathit}{cmr}{bx}{it}

  \newmathalphabet{\mathbfit} 

  \addtoversion{normal}{\mathbfit}{cmr}{bx}{it}

  \addtoversion{bold}{\mathbfit}{cmr}{bx}{it}

  \newmathalphabet{\mathbfss} 

  \addtoversion{normal}{\mathbfss}{cmss}{bx}{n}

  \addtoversion{bold}{\mathbfss}{cmss}{bx}{n}

  \ifAMStwofonts

    \ifCUPmtlplainloaded \else

      \UseAMStwoboldmath

      \makeatletter

      \new@mathgroup\upmath@group

      \define@mathgroup\mv@normal\upmath@group{eur}{m}{n}

      \define@mathgroup\mv@bold\upmath@group{eur}{b}{n}

      \edef\UPM{\hexnumber\upmath@group}

      \new@mathgroup\amsa@group

      \define@mathgroup\mv@normal\amsa@group{msa}{m}{n}

      \define@mathgroup\mv@bold\amsa@group{msa}{m}{n}

      \edef\AMSa{\hexnumber\amsa@group}

      \makeatother

      \mathchardef\upi="0\UPM19

      \mathchardef\umu="0\UPM16

      \mathchardef\upartial="0\UPM40

      \mathchardef\leqslant="3\AMSa36

      \mathchardef\geqslant="3\AMSa3E

      \let\leq=\leqslant 

      \let\geq=\geqslant

    \fi

  \fi

\fi 

\ifnfsstwo

  \DeclareMathAlphabet{\mathbfit}{OT1}{cmr}{bx}{it}

  \SetMathAlphabet\mathbfit{bold}{OT1}{cmr}{bx}{it}

  \DeclareMathAlphabet{\mathbfss}{OT1}{cmss}{bx}{n}

  \SetMathAlphabet\mathbfss{bold}{OT1}{cmss}{bx}{n}

  \ifAMStwofonts

    \ifCUPmtlplainloaded \else

      \DeclareSymbolFont{UPM}{U}{eur}{m}{n}

      \SetSymbolFont{UPM}{bold}{U}{eur}{b}{n}

      \DeclareSymbolFont{AMSa}{U}{msa}{m}{n}

      \DeclareMathSymbol{\upi}{0}{UPM}{"19}

      \DeclareMathSymbol{\umu}{0}{UPM}{"16}

      \DeclareMathSymbol{\upartial}{0}{UPM}{"40}

      \DeclareMathSymbol{\leqslant}{3}{AMSa}{"36}

      \DeclareMathSymbol{\geqslant}{3}{AMSa}{"3E}

      \let\leq=\leqslant 

      \let\geq=\geqslant

    \fi

  \fi

\fi 

\ifCUPmtlplainloaded \else

  \ifAMStwofonts \else 

    \def\upi{\pi}

    \def\umu{\mu}

    \def\upartial{\partial}

  \fi

\fi

\def\lsimeq
{\hbox{\raise0.5ex\hbox{$<\lower1.06ex\hbox{$\kern-1.07em{\sim}$}$}}}

\title[ROSAT variability of EMSS AGN ]
{ The X$-$ray flux and spectral variability of EMSS AGN}
\author[P.Ciliegi and T. Maccacaro]
       {P.Ciliegi,$^{1}$ and T. Maccacaro,$^2$ \\
1. Institute of Astronomy, Madingley Road, Cambridge CB3 OHA\\
2. Osservatorio Astronomico di Brera, Via Brera 28, 20121 Milano, Italy\\ }

\date{Accepted August 1997. Submitted December 1996}

\pubyear{1997}

\begin{document}

\maketitle

\begin{abstract}

Fifteen $ROSAT$ PSPC observations available in the public archive are 
analyzed in order to study time and spectral variability of the 12 EMSS 
AGN detected by $ROSAT$ with more than 2000 net counts. Time variability
was investigated on 13 different time scales, ranging from 400 s 
to 3.15$\times10^6$ s (1 year). Of the 12 sources analyzed, only two do not 
show a significant variability on any time scale. 
On short time scale $\leq$ 20 percent of AGN are seen as
variable sources while on time scale of $\geq$100.000 s the fraction 
becomes $\geq$50 percent. 
However one should bare in mind that the visibility function for
variability is far from being uniform and that small
amplitude variations can be detected more often on long time scale 
than on short time scale.
Spectral variability was detected in only two sources. 
MS1158.6$-$0323 shows an 
hardening of the spectrum with increasing intensity while 
MS2254.9$-$3712 shows a softening of the spectrum with increasing intensity.
Finally, for one source (MS1416.3$-$1257),
the observed variability is not 
due to an intrinsic flux variation but, instead, to a variation in the column density along 
the line of sight. Since this variability has been observed on a time scale of 
$\sim$ 3.9 days, it is probably associated to the broad line clouds. 

\end{abstract}


\section{INTRODUCTION}

\begin{figure*}

\hspace{-2.3cm}\psfig{figure=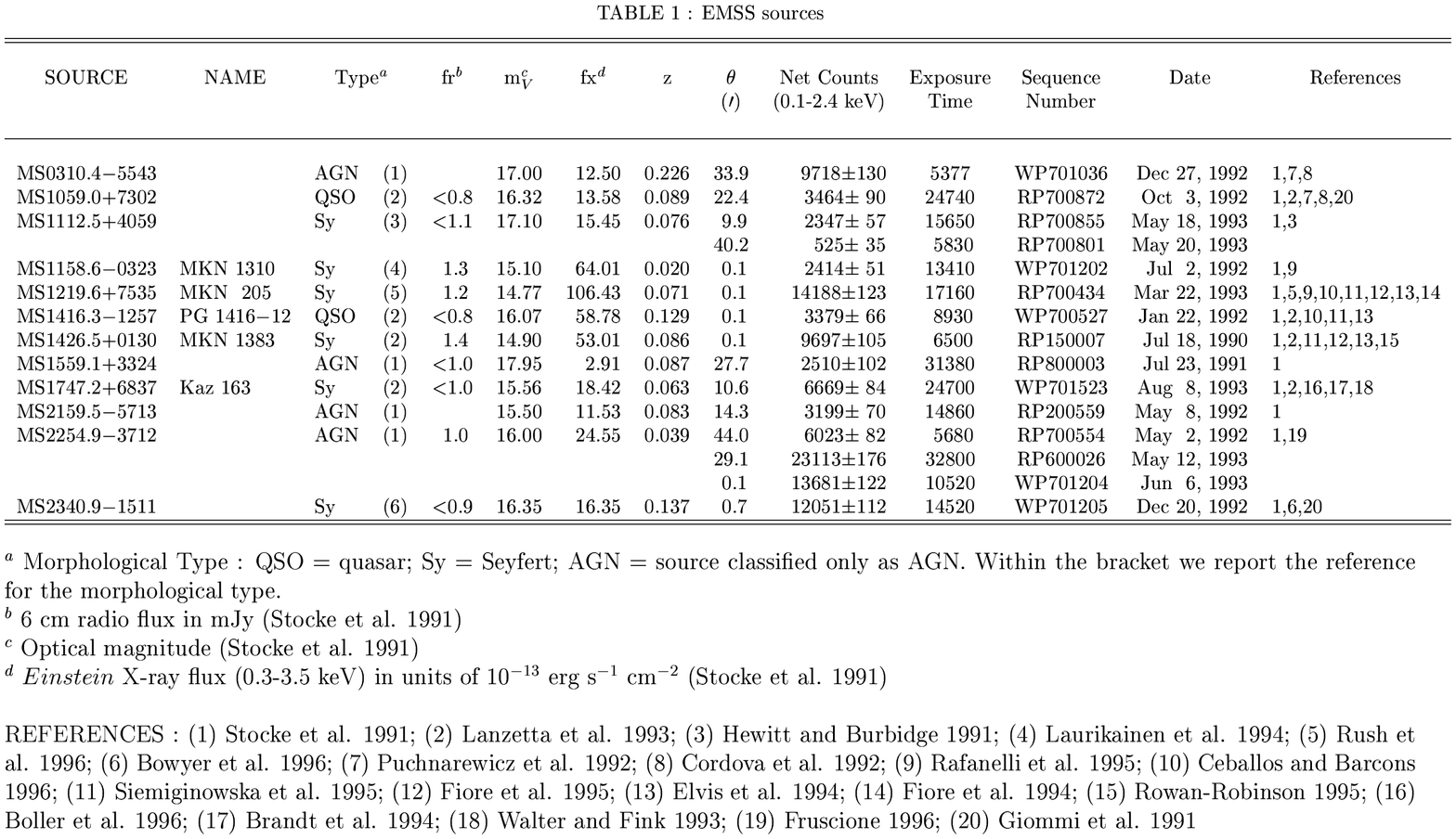,width=20cm,angle=90}
\end{figure*}

Active Galactic Nuclei (AGN) emit enormous energies over the electro magnetic 
spectrum bands from radio through gamma rays. In view of their short time 
variability, it has never been questioned that the power 
supply is primarily gravitational. The shortest
time scales  variability is seen at high energy (McHardy 1990, 
Grandi et al. 1992). Since a substantial variability cannot be observed 
on a time scale shorter than the light crossing time 
of the source (R$<$c$\delta$t), the X-ray variability is a powerful probe 
of the innermost regions of AGN and can be used to constrain the emission 
mechanisms, the size of the emitting region  and the efficiency of the 
matter/radiation conversion processes. 
 
Long duration, uninterrupted X-ray observations of AGN (hereafter with the
term AGN we refer to QSO and Seyfert 
Galaxies, i.e. {\it emission lines} AGN) 
were first carried out
by EXOSAT (White and Peacock 1989), which showed that AGN are 
X-ray sources with flux changes over a wide range of time scales and that 
short term X-ray variability, down to a few hundred seconds, is common
(McHardy 1990).  A reanalysis of the EXOSAT data (Grandi et al. 1992) 
showed that $\sim$40\% of AGN show variability on time scales less than one 
day. On longer time scales (typically weeks to months) 97\% of the same 
sample showed variability, suggesting than the long time variability is 
much more common (see also Mushotzky,
Done \& Pounds 1993). 

The EXOSAT and $Ginga$ observations have also made evident that spectral shape 
changes can accompany luminosity variations. Although the prevailing trend appears 
to be a softening of the spectrum with increasing intensity,  AGN 
do not show a unique spectral behavior (Grandi et al. 1992). This is in contrast 
to the results obtained for BL Lac objects, a particular class of AGN which 
show peculiar characteristics as a featureless optical spectrum, 
high polarization and rapid variability. Using the EXOSAT data, Giommi 
et al. (1990) have found that BL Lac show a  systematic 
hardening of the spectrum as the sources brightens. 
 
Recent $ROSAT$ observations have confirmed that the overall picture of the 
spectral variability is still rather confused. For example Boller et al. 
(1996) using a sample of narrow-line Seyfert 1 galaxies have found that 
some objects ($e.g.$ MKN 957) show a softening of the spectrum with increasing
intensity, while other sources ($e.g.$ I ZW1, MKN 507) do not show spectral 
variability associated with flux variations. Moreover, Molendi \& Maccacaro
(1994) using $ROSAT$ data on MKN 766 have found a complex spectral 
variability in which 
the 0.1$-$0.9 keV part of the spectrum hardens as the source brightens while
the 0.9$-$2.0 keV part of the spectrum  does not change significantly.
A detailed description of the expected X-ray variability in the framework 
of different theoretical models can be found in a recent paper of 
Haardt, Maraschi and Ghisellini (1997). 

Using the enormous quantity of data available in the public $ROSAT$ archive, 
we started a program aimed at the study of the X-ray properties of 
X-ray selected AGN.  In particular we are using the sample of AGN extracted 
from the $Einstein$ Extended Medium Sensitivity Survey (Gioia et al. 1990, 
Stocke et al. 1991, Maccacaro et al. 1994) for which the original 
available information
on the X-ray properties is limited by the poor energy resolution of the 
IPC and by the limited statistics of the detected sources. In the first 
paper (Ciliegi \& Maccacaro 1996) we have studied the 
X-ray spectral properties of all the EMSS AGN detected with more than 300 
net counts in $ROSAT$ PSPC images available from the public archive.  
Here we complete our work reporting the flux and spectral 
variability analysis of the EMSS AGN. In Section 2 we 
define the sub-sample of EMSS AGN used, while in Section 3 we describe the 
method of analysis. In Section 4  we report the results 
for time and spectral variability, while in Section 5 the conclusion
are presented.

\section{THE SAMPLE}

From the sample of Ciliegi \& Maccacaro (1996) of 63 EMSS AGN 
detected by $ROSAT$ with more than 300 measured net counts, we have 
extracted all the sources detected with more than 2000 
net counts. Of the 14 sources satisfying this criteria, two 
(MS0919.3+5133 and MS1215.9+3005) were 
excluded from the sample: MS0919.3+5133 was detected near 
the window support structure, so unreal flux variations may be detected 
due to the $ROSAT$ wobble and 
MS1215.9+3005 (MKN 766) was already studied in detail by Molendi et al. (1993) and
Molendi \& Maccacaro (1994)

\begin{figure*}

\psfig{figure=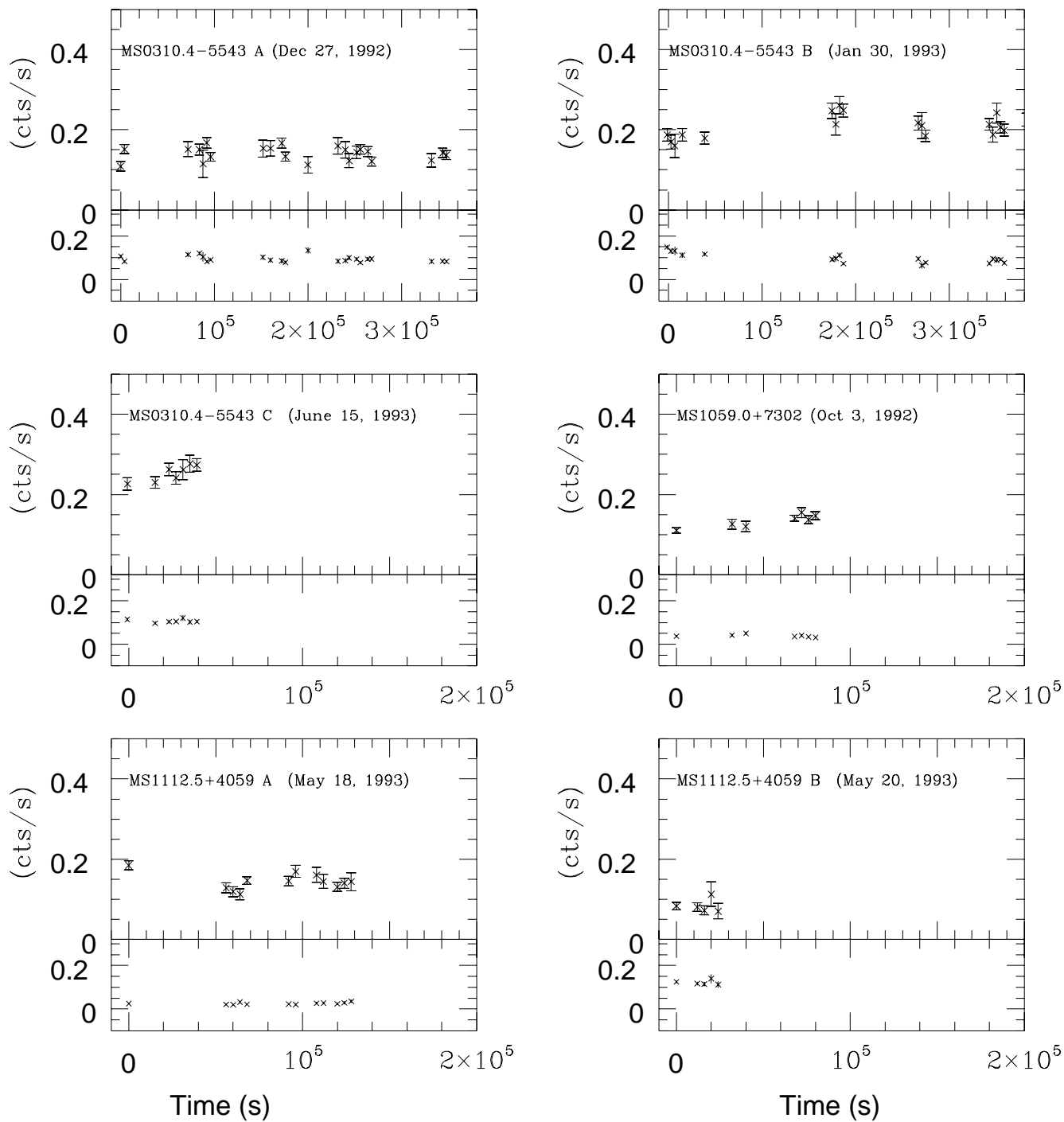,height=20cm}
\caption{Light curves for all the 12 EMSS AGN analyzed.
For each observations we report the source light curve obtained with a 
bin of 3600 s (top panel) and, for comparison, the background light 
curve (lower panel)}

\end{figure*}

\begin{figure*}
\addtocounter{figure}{-1}
\psfig{figure=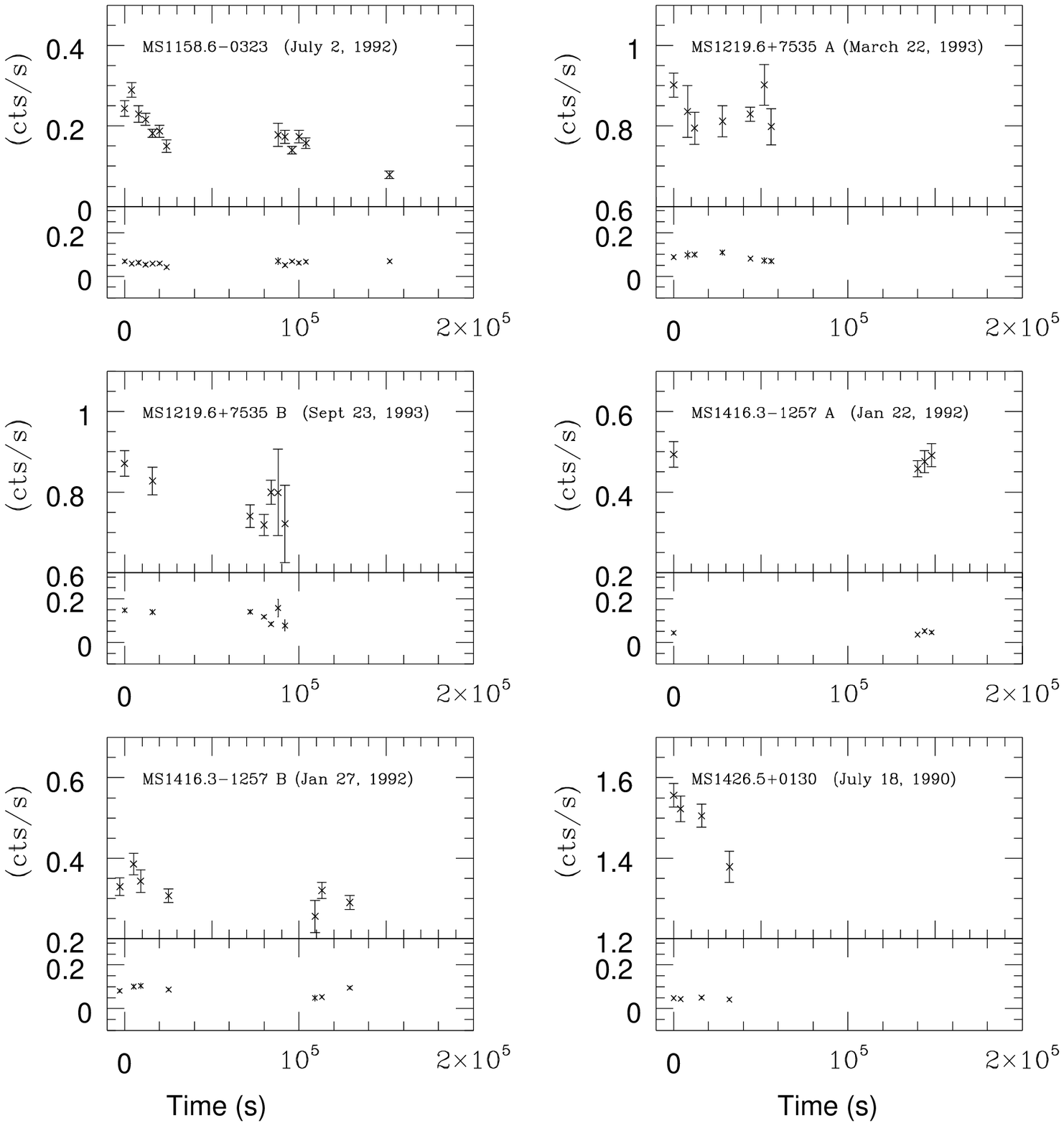,height=20cm}
\caption{continued}
\end{figure*}

\begin{figure*}
\psfig{figure=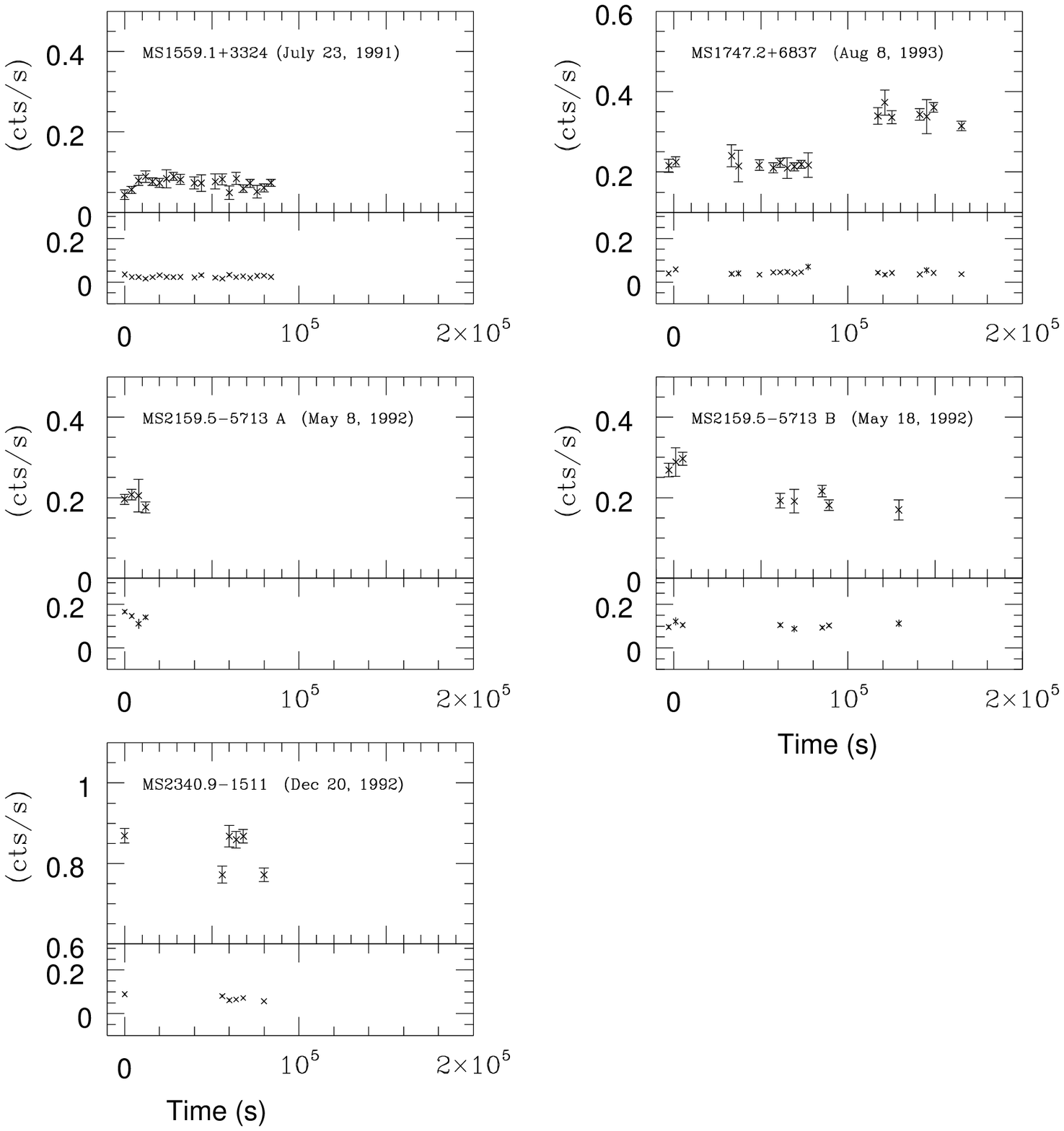,height=20cm}
\addtocounter{figure}{-1}
\caption{continued}
\end{figure*}

\begin{figure*}
\addtocounter{figure}{-1}
\psfig{figure=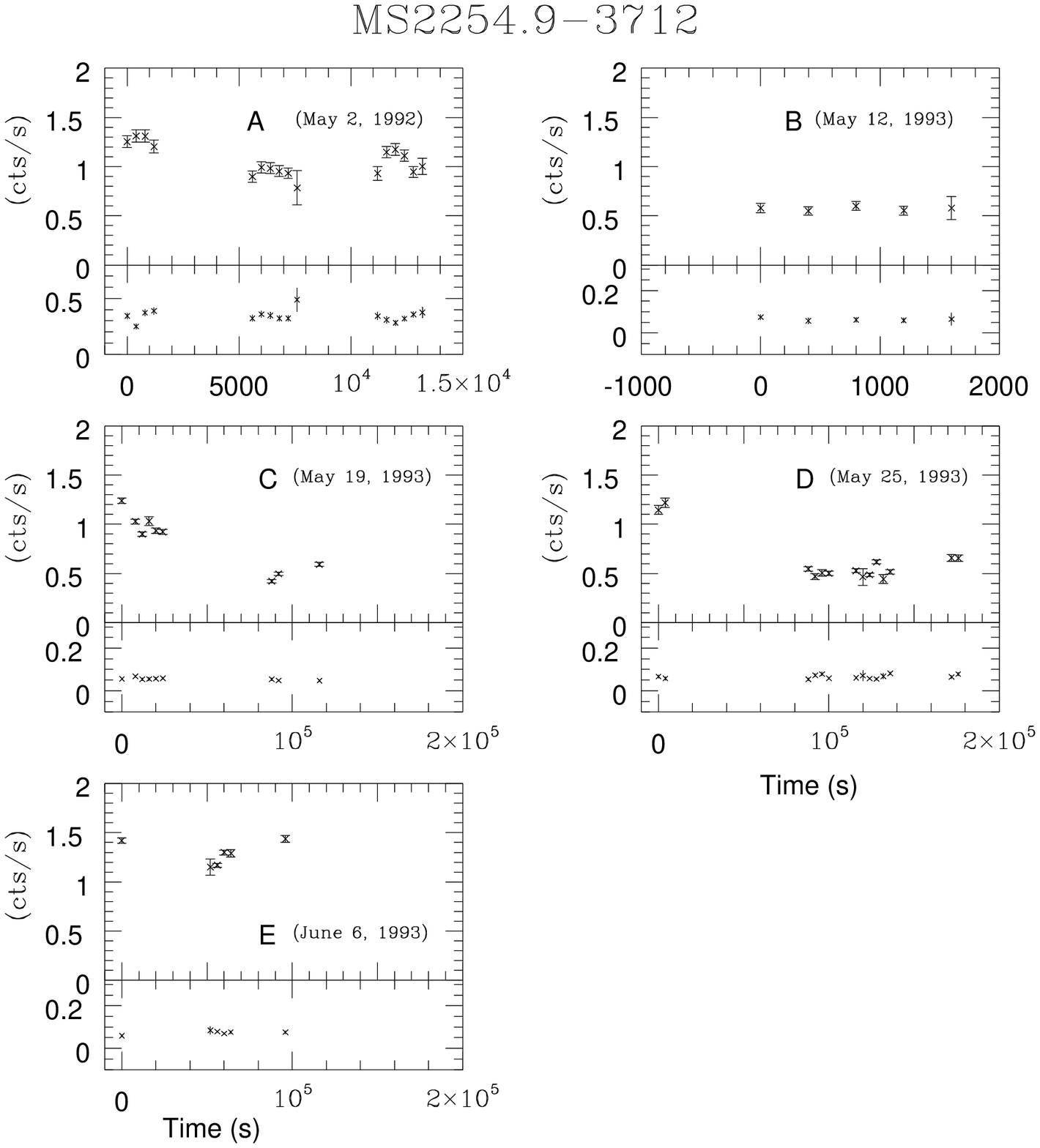,height=20cm}
\caption{continued}
\end{figure*}
 
The present sample includes 12 EMSS AGN and is listed in Table 
1 which is organized as follows: source name and morphological type 
(with relative reference within bracket) followed by 
radio flux, optical magnitude, $Einstein$ X-ray flux, redshift, off-axis
angle $\theta$ (the distance between the source and the center of the 
PSPC field), net counts detected in the 0.1$-$2.4 keV band with the 
relative error, exposure time, sequence number (ror) and 
date of the $ROSAT$ observations. Finally in the last column we report a list
(incomplete) of references to previous works. 
 We note that for MS1112.5+4059 and MS2254.9-3712
more than one observation is available.

Due to the selection criteria  
($\geq$ 2000 net counts) the 12 EMSS sources analyzed here
have optical and x-ray fluxes greater than the average of
the EMSS sample ($<$m$_V>$ = 17.9; $<$f$_x$(0.3--3.5 keV)$>$ = 7$\times 10^{-13}$ 
erg cm$^{-2}$ s$^{-1}$) 
and smaller redshifts. They 
cannot be considered representative of the whole EMSS AGN sample
but rather of its bright end.

\section{DATA ANALYSIS}

For each source we extracted the total counts in the 0.1$-$2.4 keV band
using a circle centered on the source centroid. Background counts
were estimated in an annulus centered on the source and 
subtracted from the total counts. 
A detailed description of data processing and analysis
is reported in Ciliegi \& Maccacaro (1996). 
\subsection{Time  Variability}
Time variability was investigated on 13 different time scales, ranging from 400
s to 3.15$\times10^6$ s (1 year). For all the sources we have binned the 
data in bins  that are an integer multiple of 400 s. We do this 
to avoid apparent flux variations due to the $ROSAT$ wobble period of about 
400 s. Brinkmann et al. (1994) have shown that flux determination in wobble mode 
is good to within $\sim$ 4 per cent when binning over integer multiples of 
the 400 s wobble period.  Moreover, for each time scale a careful analysis 
of the background was made to ensure that observed variability is not due 
to a change in the background rate. The light curve of all the sources obtained 
with a bin of 3600 s are shown in Figure 1. For each observation we report 
the source light curve (top panel) and, for comparison, the background light 
curve (lower panel). 
 
In order to understand and to interpret the properties of flux variability,
 attention should be paid not only to what has been detected
but also to what could or could not have been detected. The 
capability of detecting a variation of a given amplitude depends on the 
accuracy of the flux measurement, which itself depends on the total number 
of detected counts. To quantify this effect, for all the sources we have computed the 
minimum detectable variation (MDV) at 3 $\sigma$ level in each time scale  
considered.  The MDV has been obtained as follows: for a given time scale,
a variation of a fraction K in flux  
(defined as $\mid$C$_{i+1} - $C$_i\mid$/C$_i$) 
between two consecutive bins can be detected (with a significance level of N 
$\sigma$) if \\
$\mid$C$_i -$ C$_{i+1}\mid>$N$\times \sqrt{E_i^2 + E_{i+1}^2}$,
where C$_i$, C$_{i+1}$, $E_i$ and $ E_{i+1}$ are the net counts and the relative errors
in the $i$th and in the $i+1$th bin.  Writing C$_{i+1}$ = C$_i$(1+K) and 
assuming $E_{i+1}=E_i$(1+K)$^{1/2}$ we have: \\
C$_i$K $>$ N $\times \sqrt{E_i^2(1+(1+K))}$ 
which gives
$E_i$/C$_i < \sqrt{K^2/(N^2(1+(1+K))}$. \\
For example, to be able to detect a variation of 30 per cent 
(K=0.3) at a significance level of 3 $\sigma$ (N=3), we must have  
$E_1$/C$_1 <$ 0.066 in at least two consecutive bin. 
 
\subsection{Spectral variability}

For the sources that show flux variability we have calculated the 
hardness ratio HR for each bin of the light curve in order to study a possible 
correlation between HR and 
count rate. The hardness ratio provides a powerful tool for the detection 
of X-ray spectral variability, as the hardness ratio is model independent. 
The hardness ratio HR is defined as HR=(H$-$S)/(H+S) where S is the number 
of net counts detected in the 0.11$-$0.43 keV band and H is the number 
of net counts detected in the 0.50$-$2.02 keV band. 
 
Moreover, in order to carry out a less detailed but statistically more 
significant analysis, for each source we have divided the $ROSAT$ data
into two subsets depending on whether the count rate is below (low state) 
or above (high state)
the mean count rate  and  we have computed the hardness ratio 
for each subset. 
The results of the time and spectral variability analysis are reported in the next
Section.  
 
\section{Results}

For all the time scale considered we have calculated the observed variability 
(ObV) at the 3 $\sigma$ level for each of the 12 EMSS AGN detected
by $ROSAT$ with more than 2000 net counts. The observed variability, calculated 
between all the consecutive bins, is defined as 
ObV = $\mid$CR(i+1) - CR(i)$\mid$/CR(i), where CR(i) and CR(i+1) are the 
observed count rate in the  $i$th and in the $i+1$th bin respectively. 
For each time scale, in Table 2 we report the minimum detectable variability 
(MDV), the minimum and the maximum value of the observed variability and 
(within bracket) the number of times that we detected a significant 
variability at 3 $\sigma$ level. 

As shown in the Table, not all the sources have been observed over the same time 
scales. This is due to the fact that the sampling of the data is 
fairly heterogeneous. For some sources the sampling is fairly constant over 
few days, while for other sources is strongly discontinuous with gaps 
of week or months ($i.e.$ time that the satellite has spent on other sources). 
We note that over long time scales the detected variability results from the comparison
of observations separated by a significant amount of time during which
the source was not monitored (see for instance the case of MS2254.9$-$3712, 
observed on May 2, 1992 and on May 12, 1993). It is clear that the statement
that the source has varied over a time scale of one year is true but the
one year time scale should
be considered as an upper limit. We do not know whether the transition from a lower
to a higher count rate occurred smoothly over several months or abruptly in
a much shorter time scale.

Of the 12 sources analyzed, only two (MS1059.0+7302 and MS1559.1+3324) do not 
show a significant variability on any time scale. In Figure 2 we report 
the fraction of the 

\begin{figure}
\psfig{figure=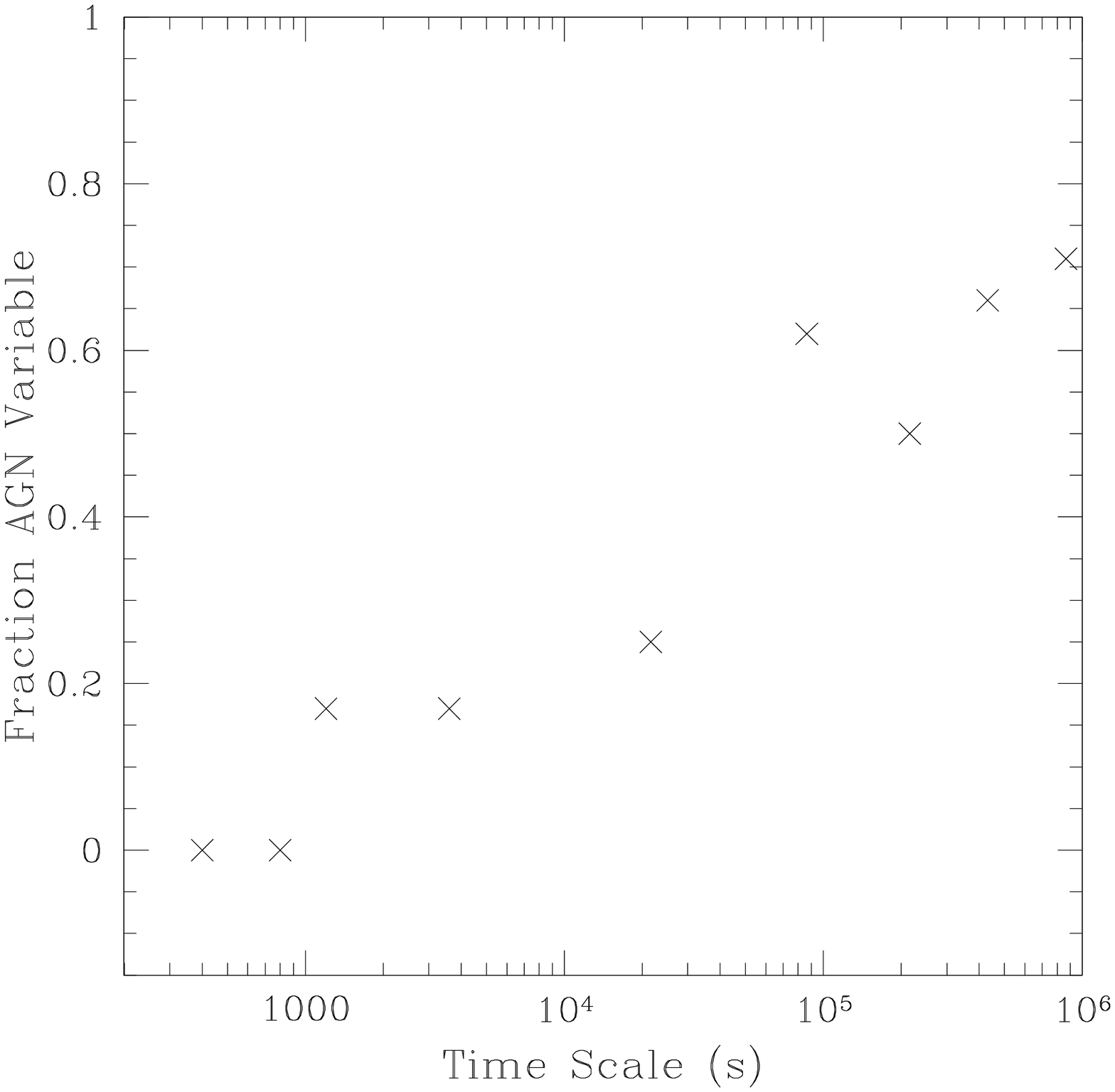,width=8cm}
\caption{ The fraction of the sources that show a significant variability
as a function of time scale }
\end{figure}

sources that show a significant variability as a function 
of the time scale. As clearly shown in Figure 2, long time scale variability 
is much more common than short time scale variability, in agreement with the 
results obtained by Grandi et al. (1992) with the EXOSAT data. 
Table 2 also shows that the typical amplitude of the variability 
observed on short time scales are, in general, lower than that on long time 
scales. With few exceptions, the amplitude of the variation on time scale 
less than 30 hours are always lower than 30 per cent, while on longer time 
scale the amplitude of the variations are always greater than 30 per cent. 
Therefore the increasing fraction of variable source on longer time scale 
is due to the combination of two effects: an increase in the amplitude 
of the variation and the contemporaneous decrease of the MDV due to the 
better statistic available on longer time scales. 

\begin{figure*}
\hspace{-2.5cm}\psfig{figure=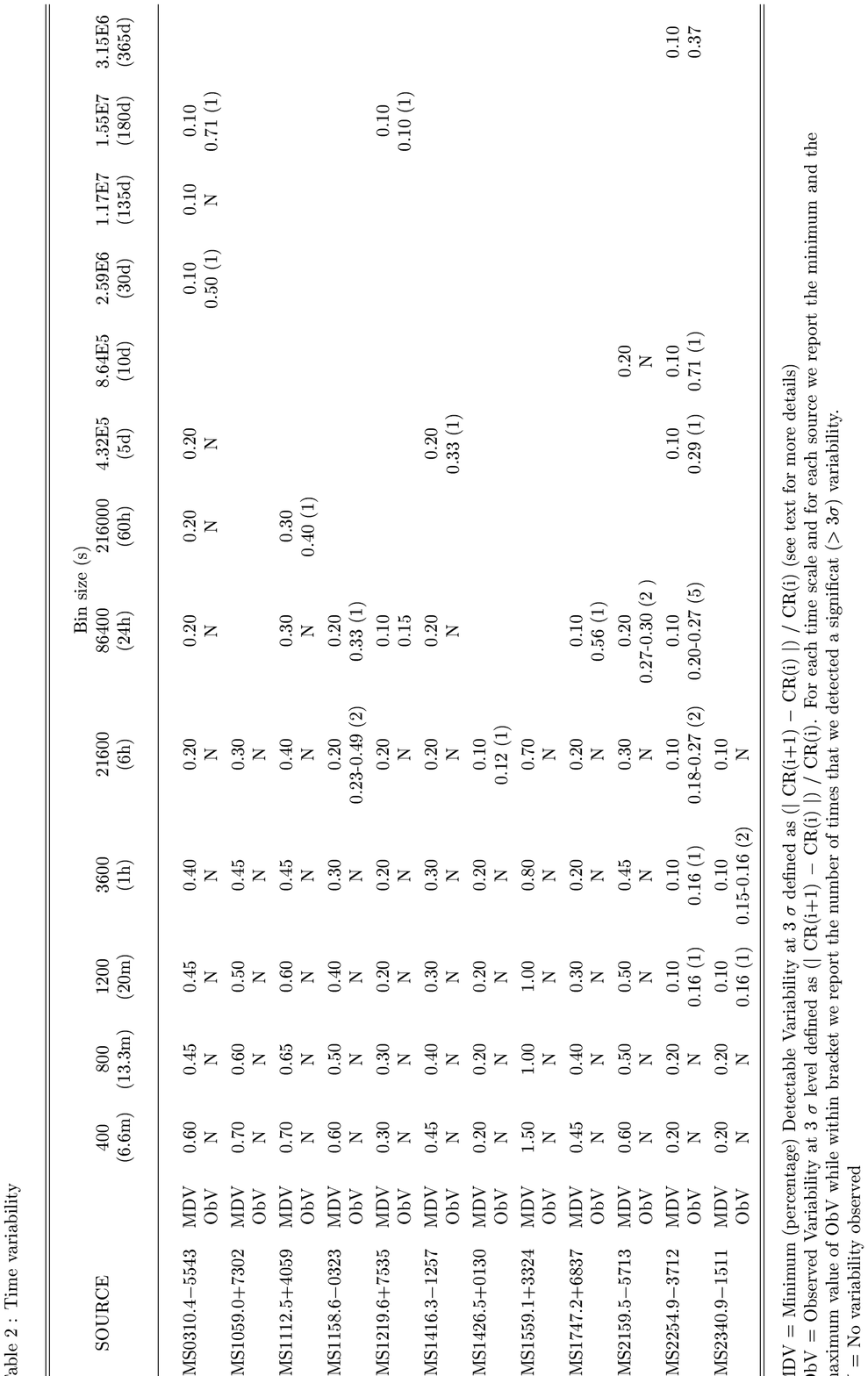,angle=180,height=24cm}
\end{figure*}

For the 10 sources that show a significant variability, we also searched for 
the shortest time scale on which a significant variability  occurs. The 10 
variable sources are listed in Table 3 where we report the source name, 
the shortest time scale on which we detected a significant ($>3\sigma$) 
variability, the percentage flux variation occurred during the shortest 
time scale (defined as in Table 2) and the corresponding luminosity variation. 
Assuming that the luminosity is produced by matter being transformed into 
radiation with some efficiency $\eta$, we have calculate $\eta$ for each 
source using $\Delta L \leq (\eta$ m$_p$ c$^4$ t$_{var}$) / $\sigma_T$ 
where $\sigma_T$ is the Thomson cross-section and m$_p$ the mass of proton 
(Fabian 1992). For all the sources the observed luminosity variation imply 
an efficiency $\eta <$ 0.1, so there is no indication that the observed 
X-ray radiation is beamed. 

Of the 10 sources with detected flux variability, only 3 show 
also a spectral variation associated with the flux variation. Below 
we discuss the flux and spectral properties of these three sources. 

\begin{table}
\begin{center}
\footnotesize
 
TABLE 3 : Sources with detected  variability  \\
 
\vspace{3mm}
 
\begin{tabular}{lrcc} \hline \hline
 &  &  \\
\multicolumn{1}{c}{SOURCE} & \multicolumn{1}{c}{$\Delta T_{\tiny \rm min}$} 
& $\Delta F$ & $\Delta L$ \\
 & \multicolumn{1}{c}{(s)} & & erg s$^{-1}$ \\
 &    & \\ \hline
 &    & \\
         MS0310.4$-$5543     & 2.56$\times 10^{6}$ & 50\% & 4.02$\times 10^{45}$  \\[2.5mm]
         MS1112.5+4059       & 2.16$\times 10^{5}$ & 40\% & 2.86$\times 10^{43}$  \\[2.5mm]
         MS1158.6$-$0323     &  8000               & 32\% & 1.17$\times 10^{42}$  \\[2.5mm]
         MS1219.6+7535       & 72000               & 15\% & 1.04$\times 10^{44}$  \\[2.5mm]
         MS1416.3$-$1257     & 4.32$\times 10^{5}$ & 33\% & 3.07$\times 10^{44}$  \\[2.5mm]  
         MS1426.5+0130       & 14400               & 17\% & 2.01$\times 10^{44}$  \\[2.5mm]
         MS1747.2+6837       & 40000               & 56\% & 1.59$\times 10^{44}$  \\[2.5mm]
         MS2159.5$-$5713     & 56000               & 33\% & 7.03$\times 10^{43}$  \\[2.5mm]
         MS2254.9$-$3712     &  1200               & 16\% & 1.25$\times 10^{43}$  \\[2.5mm]
         MS2340.9$-$1511     &  1200               & 16\% & 2.39$\times 10^{44}$  \\[2.5mm] 
\hline \hline
\end{tabular}
 
\end{center}
 
\end{table}

\subsection {Sources which show flux and spectral variability}

\noindent {\bf MS1158.6$-$0323} (MKN 1312)

For this source 
we detected an X-ray count rate decrease of $\sim$32 per cent
 on a time scale of 
8000 seconds, while over the whole $ROSAT$ observation the ratio of maximum 
to minimum count rate was $\sim$ 3.6. We searched for spectral variability 
calculating the hardness ratio in each bin of the light curve. In Figure 3 
we report the 3600 s 
light curve of MS1158.6$-$0323 and, for comparison, the value 
of the hardness ratio in each bin. Figure 3 shows 
an hardening of the spectrum as the source brightens, although this trend 
is not statistically significant (the error bars represent the 
1 $\sigma$ error).   
However, if we divide the data into two subsets from 0.08 to 0.18 
counts s$^{-1}$ (low state) and from  0.18 to 0.29 counts s$^{-1}$ (high state)
the hardening of the spectrum becomes significant, 
with HR$_{\em LOW~STATE}=0.26\pm0.02$ and 
HR$_{\em HIGH~STATE}=0.34\pm0.02$. \\

\begin{figure}
\psfig{figure=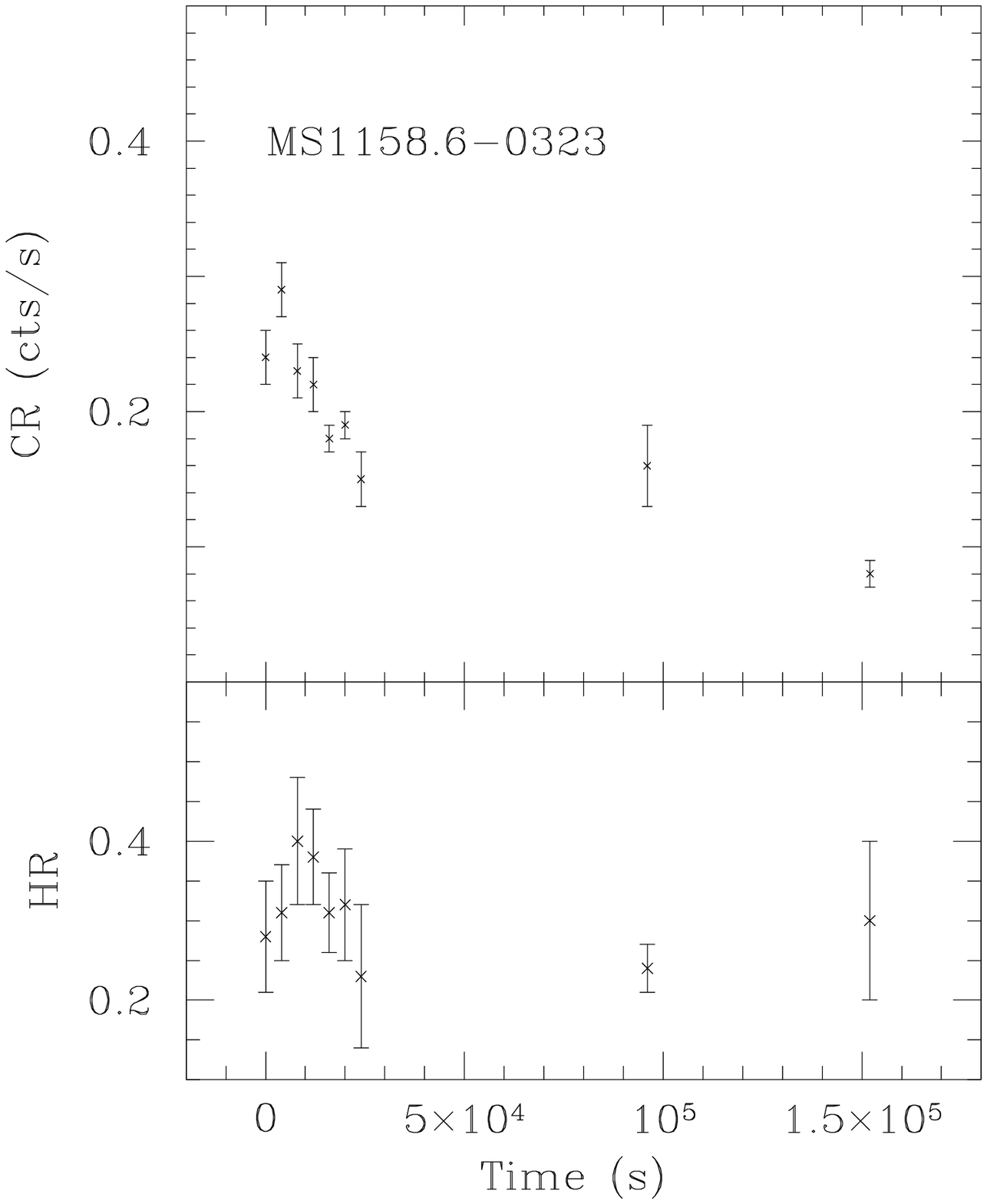,width=8cm}
\caption{ The light curve of MS1158.6$-$0323 obtained with a bin of 
3600 s (top panel) and the value of the hardness ratio HR in each bin 
(lower panel)}
\end{figure}

\noindent {\bf MS1416.3$-$1257} (PG1416-12)

This source show a significant X-ray count rate decrease of 33 per 
cent on a time scale of 4.32$\times10^5$ s ($\sim$ 5 days). 
Therefore we splitted the observation of MS1416.3$-$1257 into two data sets 
separated by 4.32$\times10^5$ s and we searched for spectral 
variability. The spectrum 
in the first data set is well parameterized by fitting the data with a 
broken power-law with N$_H$ fixed at the Galactic value (6.8$\times10^{20}$ 
cm$^{-2}$). It yields $\alpha_x$=1.20$\pm$0.06 and $\chi^2$/dof=20.55/26 with
a flux (unabsorbed for Galactic N$_H$) of 118.3$\pm$3.32 10$^{-13}$ erg 
cm$^{-2}$ s$^{-1}$ (0.1$-$2.4 keV band). Modeling the data with N$_H$ as a
free parameter does not yield a significant improvement to the fit, with 
$\alpha_x$=1.18$\pm$0.21, N$_H$=6.69$^{+1.02}_{-0.93} \times 10^{20}$ cm$^{-2}$
and $\chi^2$/dof=20.53/25. 

In the second data set, the spectrum shows excess 
absorption relative to the Galactic value. In fact, while the model with 
N$_H$=N$_{H~Gal}$ yields $\alpha_x=1.00\pm0.11$ and $\chi^2$/dof=30.35/21, 
the model with N$_H$ free to vary yields $\alpha_x=1.35\pm0.23$, 
N$_H$=9.7$\pm1.4\times10^{20}$ cm$^{-2}$ and $\chi^2$/dof=17.27/20. The 
reduction in $\chi^2_{FIT}$ with the addition of N$_H$ as a free 
parameter is statistically significant (P$_{F>F_{FIT}}$=9.0 10$^{-4}$ 
with F test). The absorption excess can be well parameterized by fitting the 
data with a power-law ($\alpha_x=1.35\pm$0.25) absorbed by Galactic N$_H$ 
plus a second absorption component (N$_H$=2.86$^{+2.19}_{-0.75} \times
10^{20}$ cm$^{-2}$)  at the redshift of the source (z=0.129). Using this 
model, the flux of the source, unabsorbed for Galactic N$_H$ and for 
the second absorption component, is f$_x$(0.1$-$2.4)=121.3$\pm$3.32 
10$^{-13}$ erg cm$^{-2}$ s$^{-1}$, consistent with the flux observed in the 
first observation. Therefore the observed variability of MS1416.3$-$1257
is not due to a flux variation of the source, but to a variation in the 
column density. 

Variations in the column density are not common events in AGN. They have been
seen in only few object like, for example, NGC4151 (Yaqoob et al. 1993), 
ESO103$-$G55 (Warwick et al. 1988), NGC6814 (Leighly et al. 1992), 
NGC5506 (Bond et al. 1992). The presence of a significant amount of X-ray 
column density in AGN is generally associated with the accretion disk, the 
cold torus, the disk of the galaxy itself (Lawrence and Elvis 1982) or with 
the broad line clouds. The apparent rarity of variations in absorption is 
consistent with all these explanations, although the occurrence of 
variability at all argues for a component associated with either the 
broad line clouds or the edge of the accretion disk, since the other two 
regions are not expected to change on observable time scales.

\noindent {\bf MS2254.9$-$3712}

\begin{figure}
\psfig{figure=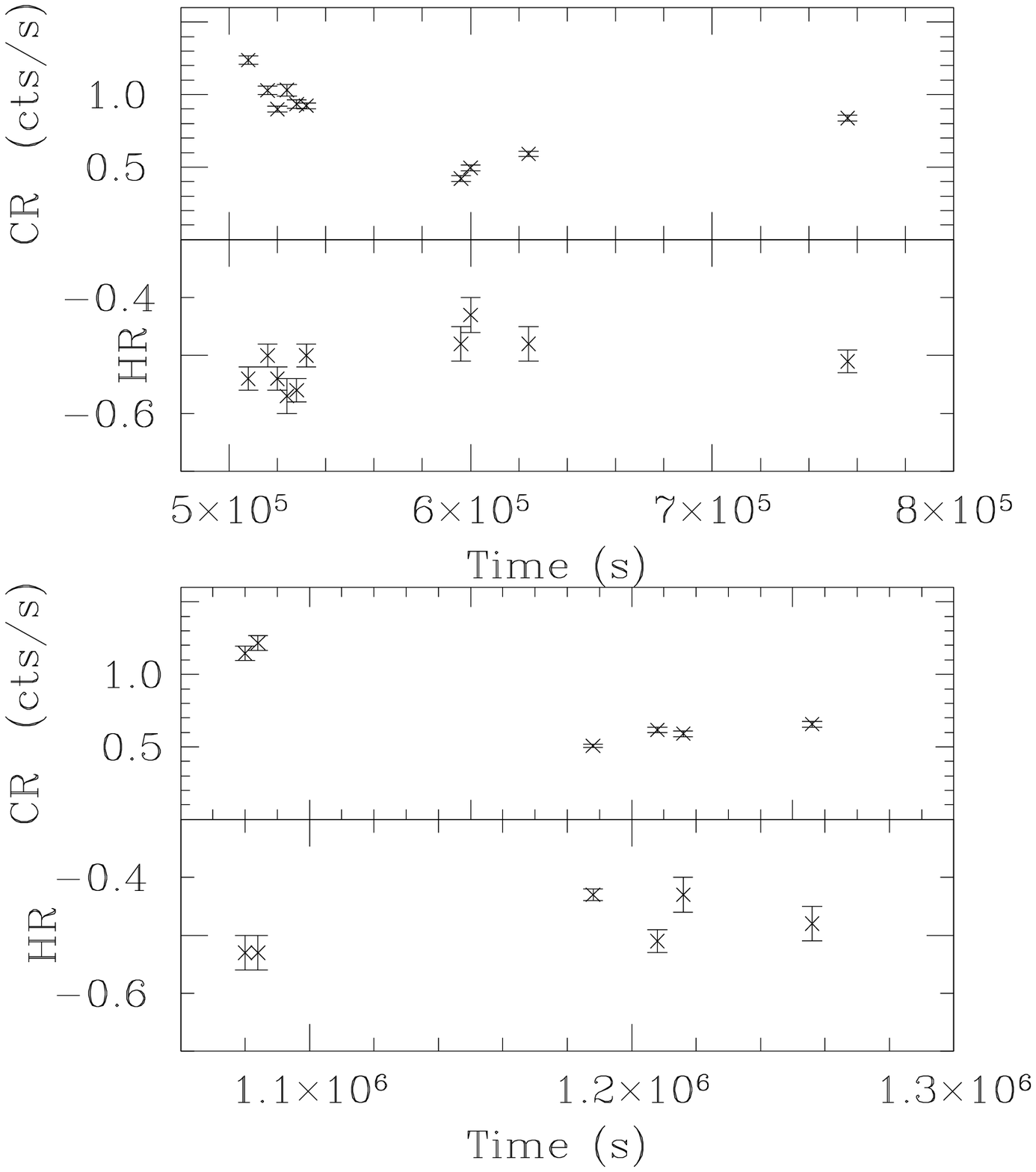,width=8cm}
\caption{ As in Figure 3 for MS2254.9$-$3712}
\end{figure}

For this source we have 3 different observations (see Table 1). 
As shown in Figure 1, 
this source show strong variability both on short and long time scales. A significant 
increase and decrease of the count rate of $\sim$16 per cent were detected on time 
scales of 1200 s, while on a time scale of 86800 s the 
source count rate decreased by a factor of $\sim$3.3. 

We searched for spectral variability calculating the hardness ratio in each bin of 
the light curve. The results are presented in Figure 4 which shows a complex 
pattern, although the prevailing 
trend appears to be a softening of the spectrum with increasing intensity.
This is confirmed by Figure 5 where we report the hardness ratio HR as a function 
of count rate. 
If we  divide the data into two subset: from 0.51 to 0.90 counts s$^{-1}$ (low state) 
and from 0.90 to 1.43 counts s$^{-1}$ (high state), the softening of the spectrum 
with increasing
intensity becomes significant, with HR$_{\em LOW~STATE}=-0.46\pm0.01$ and 
HR$_{\em HIGH~STATE}=-0.53\pm0.01$

\begin{figure}
\psfig{figure=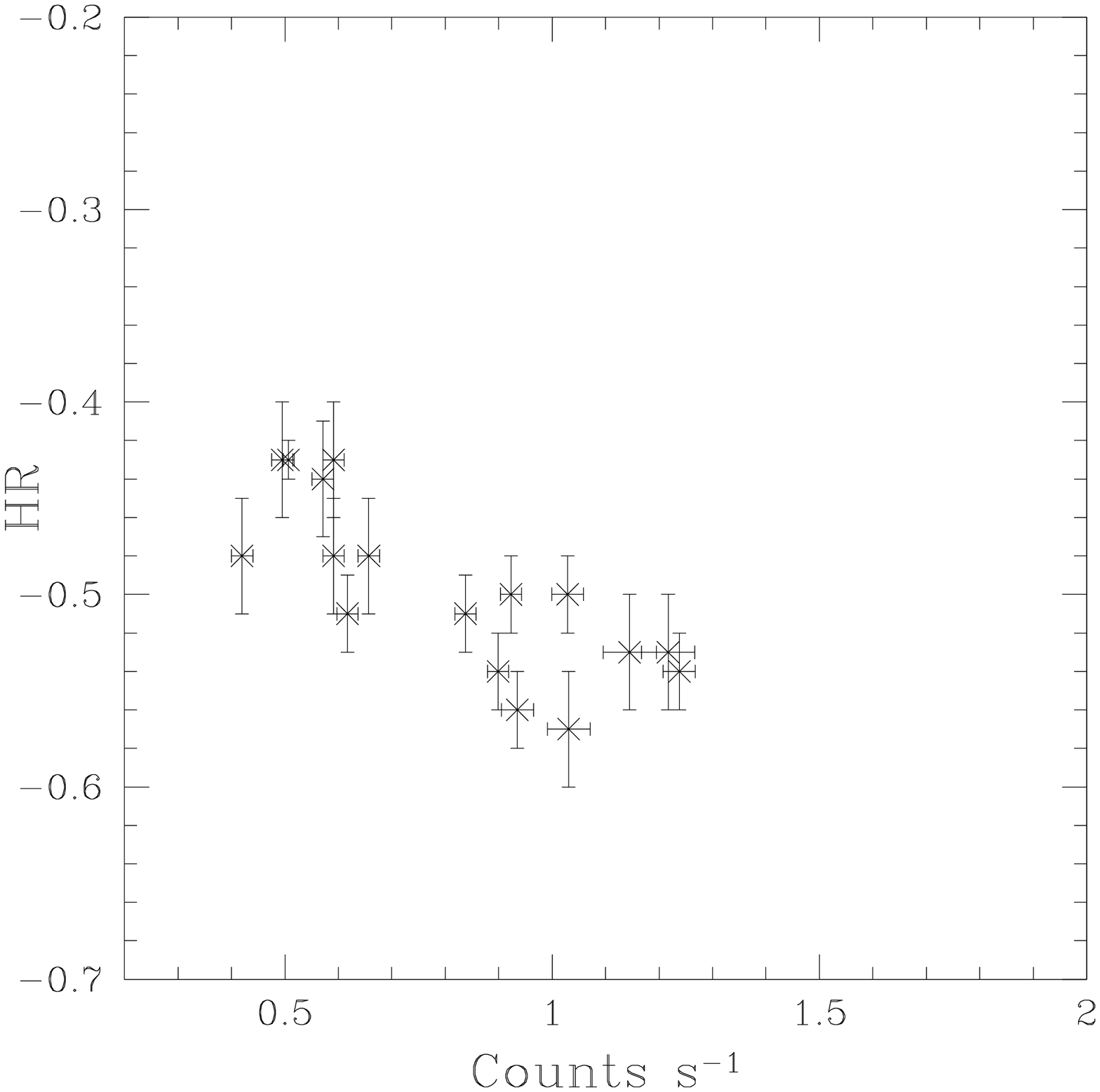,width=8cm}
\caption{ The hardness ratio HR as a function of count rate for 
MS2254.9$-$3712}
\end{figure}

\section {Conclusions}

The flux and spectral variability analysis of a sample of 12 EMSS AGN has confirmed 
that long term ($\geq$ 30 hours) variability is more common than short term 
variability. However this effect is probably due to a selection effect more than
to a real increase of the 
fraction of AGN that show a variability. In this respect we note that Fig. 2
shows an increase of the fraction of AGN exhibiting flux variation with increasing
time scale. On short time scale about 20 percent or less of AGN are seen as
variable sources while on time scale of 100.000 s or more the fraction becomes $\geq$
50 percent. However one should bare in mind that the visibility function for
variability is far from being uniform. Inspection of table 2 shows that small
amplitude variations can be detected more often on long time scale than on short time
scale.

On the other end we note that, considering only sources for which variability
has been detected, on time scale less than 30 hours the amplitude 
of the variation are mainly lower than 30 per cent while on longer time scale 
the amplitude are mainly greater than 30 per cent. 

Unfortunally, using only the ROSAT 
data we are not able to determine the very nature of the detected variability 
and to understand if the different intensity detected on time scale shorter and longer
than 30 hours is due to different physical and/or geometrical properties. 

As shown by Haardt, Maraschi and Ghisellini (1997), the only way to determine the 
real nature of an observed X-ray variability is an accurate analysis of the flux 
variability as a function of the spectral variability in the 2$-$10 keV band 
together with simultaneous observation in the 0.1$-$2.0 keV band. 

However, it is interesting to note that a time scale of 30 hours corresponds to an 
upper limit to the size of the source (R$\leq$c$\delta$t) of $\sim$ 
3$\times 10^{15}$ cm.  In the most accepted picture of the physical structure 
of AGN (see, for example, Urry and Padovani 1995), this region is between 
the accretion disk around the central black hole ($\sim1-30\times 10^{14}$ cm
for a central black hole of 10$^8$M$_{\odot}$) and the region of the broad line 
clouds ($\sim2-20\times 10^{16}$ cm). Therefore, while the variability on time 
scale shorten than 30 hours is surely associated with a variation in the 
accretion disk, on longer time scale it can arise from larger regions where 
other factors ($i.e.$ scattered radiation, absorbing clouds) can contribute
to the observed variability. 

For one source  (MS1416.3$-$1257) the observed variability is not 
due to a flux variation but, instead, to a variation in the column density along 
the line of sight. Since this variability has been observed on a time scale of 
$\sim$ 3.9 days, it is probably associated to the broad line clouds. 

Spectral variability was detected in only two sources. MS1158.6$-$0323 shows an 
hardening of the spectrum with increasing intensity while MS2254.9$-$3712 shows 
a softening of the spectrum with increasing intensity. Finally 
MS1215.9+3005 (MKN766), which was excluded from our analysis because already 
studied in detail by Molendi and Maccacaro (1994) but is part of the selected sample, 
shows a different
behavior between the soft (0.1-0.9 keV) and hard (0.9-2.0 keV) part of 
the spectrum. The soft part harden as the source brightens while the hard 
part does not change significantly. These results confirm  that the 
overall picture of the flux and spectral variability is still rather confused.  
An improvement of our knowledge in this field should come from the data of the 
recently launched XTE and SAX X-ray satellite and from the forthcoming AXAF and
XMM missions, characterized by the possibility of long, uninterrupted observations
over a broader energy band.  

\section*{Acknowledgments}

We thank the referee for useful comments and criticisms. This work has received 
partial financial support from the Italian Space Agency
(ASI contract 95-RS-72 161FAE/2).
This research has made use of the NASA/IPAC Extragalactic Database (NED)   
which is operated by the Jet Propulsion Laboratory, California Institute   
of Technology, under contract with the National Aeronautics and Space      
Administration.

\end{document}